\documentclass{elsart}

\usepackage{amssymb}
\usepackage{natbib}
\usepackage{dsfont}

\date{August 2004}

\begin{document}
\begin{frontmatter}

\title{Holism, Physical Theories and Quantum Mechanics\thanksref{talk}}
\thanks[talk]{This paper is based on talks given at the `International Workshop on Holism in the Philosophy 
of Physics' held at Bonn on  4-5 July 2003 and at the `12th UK Conference on the Foundations of Physics'
held at Leeds on 2-5 September 2003.}

\author[ik]{M.P. Seevinck}
\ead{m.p.seevinck@phys.uu.nl}
\address[ik]{Institute for History and Foundations of Natural Science, P.O.Box 80.000, University of Utrecht,
 3508 TA Utrecht, The Netherlands}

\begin{abstract}
Motivated by the question what it is that makes quantum mechanics a 
holistic theory (if so), I try to define for general physical theories what 
we mean by `holism'. For this purpose I propose an epistemological criterion to 
decide whether or not a physical theory is holistic, namely: a physical theory 
is holistic if and only if it is impossible in principle to infer the global 
properties, as assigned in the theory, by local resources available to an agent.  
I propose that these resources include at least all local operations and 
classical communication. This approach is contrasted with the well-known 
approaches to holism in terms of supervenience. The criterion for holism 
proposed here involves a shift in emphasis from 
ontology to epistemology. I apply this epistemological criterion to 
classical physics and Bohmian mechanics as represented on a phase and 
configuration space respectively, and for quantum mechanics (in the orthodox 
interpretation) using the formalism of general quantum operations as completely 
positive trace non-increasing maps. Furthermore, I provide an interesting 
example from which one can conclude that quantum mechanics is holistic in the 
above mentioned sense, although, perhaps surprisingly, no entanglement is needed.
\end{abstract}

\end{frontmatter}

\section{Introduction}
\label{inleiding}

Holism is often taken to be the idea that the whole is more than the sum of the parts.
Because of being too vague, this idea has only served as a guideline
or intuition to various sharper formulations of holism.
Here I shall be concerned with the one relevant to physics, i.e., the doctrine of 
\emph{metaphysical holism}, which is the idea that properties or relations of a 
whole are not determined or can not be determined by intrinsic properties or relations of the
parts\footnote{This \emph{metaphysical holism} (also called \emph{property holism})
 is to be contrasted  with \emph{explanatory} and \emph{meaning holism} \citep{healey91}.
  The first is the idea that explanation of 
a certain behavior of an object cannot be given by analyzing the component 
parts of that object. Think of consciousness of which some claim that it cannot be
 fully explained in terms of physical and chemical laws obeyed by the molecules of 
 the brain. The second  is the idea that the meaning of a term cannot be given without regarding 
it within the full context of its possible functioning and usage in a language.}.
  This is taken to be opposed to a claim of supervenience \citep{healey91},
   to reductionism \citep{maudlin98}, to local physicalism \citep{teller86}, and 
   to particularism \citep{teller89}.
   In all these cases a common approach is used to define what metaphysical holism
is: via the notion of \emph{supervenience}\footnote{The notion of supervenience, as used here, is meant to describe 
a particular relationship between properties of a whole and properties of the parts of that whole.
The main intuition behind what particular kind of relationship is meant, is  
captured by the following impossibility claim. It is not possible that two things should
 be identical with respect to their subvenient or subjacent properties (i.e., the lower-level properties),
without also being identical with respect to their  supervening or upper-level properties.   
The first are the properties of the parts, the second are those of the whole. 
The idea is that there can be no relevant difference in the whole without 
a difference in the parts. (\citet{cleland84} uses a different definition in terms of
 modal logic.)}.
\label{supervenience}
According to this common approach metaphysical holism is the doctrine
 that some facts, properties, or relations of the whole do not supervene on 
intrinsic properties and relations of the parts, the latter together making up the \emph{supervenience basis}. 
As applied to physical theories, quantum mechanics is then taken to be the paradigmatic
example of a holistic theory, since certain composite states
 (i.e., entangled states) do not supervene on subsystem states, a feature not 
 found in classical physical theories.

However, in this paper I want to critically review the supervenience
approach to holism and propose a new criterion for deciding whether or not a physical theory is holistic. 
The criterion for whether 
or not a theory is holistic proposed here is an \emph{epistemological} one. 
It incorporates the idea that each physical theory 
(possibly supplemented with a property assignment rule via an interpretation) 
has the crucial feature that it tells us how to \emph{actually} infer 
properties of systems and subsystems.

The guiding idea of the approach here suggested, is that some property
of a whole would be holistic if, according to the theory in question, there is no 
way we can find out about it using \emph{only} local means,
i.e., by using only all possible non-holistic resources available to an agent.
In this case, the parts would not allow for inferring the properties of the 
whole, not even via all possible subsystem property determinations that 
can be performed, and consequentially we would have some instantiation of holism, 
called \emph{epistemological holism}.
The set of non-holistic resources is called the \emph{resource basis}. I propose that this basis includes 
at least all local operations and classical communication of 
 the kind the theory in question allows for.
 
The approach suggested here thus focuses on property measurement 
instead of on the supervenience of properties. 
It can be viewed as a shift from ontology to 
epistemology\footnote{This difference is similar to the difference
between the two alternative
definitions of determinism. From an \emph{ontological} point of view,
determinism is the existence of a single possible
  future for every possible present.
  Alternatively, from an \emph{epistemological} point of view, it is the 
  possibility in principle of inferring the future from 
  the present.} and also as a shift that takes into account the full 
  potential of physical theories by including 
  what kind of property inferences or measurements are
   possible according to the theory in question. 
    The claim I make is that these two approaches are crucially different and that
 each have their own merits. I show the fruitfulness of the new approach 
 by illustrating 
it in classical physics, Bohmian mechanics and orthodox quantum mechanics. 
 
The structure of this paper is as follows. First I will present in section 2 a short review 
of the supervenience approach to holism. I especially look at the supervenience 
 basis used. To illustrate this approach 
 I consider what it has to say about classical physics and quantum mechanics. 
 Here I rigorously show that in this approach classical physics is non-holistic and furthermore
  that the orthodox interpretation of quantum mechanics is deemed holistic. 
  In the next section 
 (section 3) I will  give a different approach based on an epistemological 
 stance towards property determination within physical theories.
This approach is contrasted with the approach of the previous section and furthermore 
argued to be a very suitable one for addressing holism in physical theories.

In order to show its fruitfulness I will apply the epistemological approach to different physical theories. 
Indeed, in section 4 classical physics and Bohmian mechanics are 
proven not to be epistemologically holistic, whereas the orthodox interpretation 
of quantum mechanics is 
shown to be epistemologically holistic without making appeal to the feature of 
entanglement, a feature that was taken to be absolutely necessary in 
the supervenience approach for any holism to arise 
in the orthodox interpretation of quantum mechanics.
Finally in section 5 I will recapitulate, and argue this new approach to holism 
to be a fruit of the rise of the new field of quantum information theory.

\section{Supervenience approaches to holism}
\label{supervenientie-sectie}

The idea that holism in physical theories is opposed to supervenience of 
properties of the whole on intrinsic properties or relations
 of the parts, is worked out in detail by \citet{teller86} and by 
 \citet{healey91}, although others have used this idea as
  well, such as \citet{french89}\footnote{
\citet{french89} uses a slightly different approach to holism
 where supervenience is defined in terms of modal logic, following a proposal 
 by \citet{cleland84}. 
 However, for the present purposes, this approach leads essentially
 to the same results and I will not discuss it any further.}\label{voetnootfrech},
  \citet{maudlin98} and \citet{esfeld01}. I will review 
  the first two contributions in this section.

Before discussing the specific way in which part and whole are related,
\citet{healey91} clears the metaphysical ground of what 
it means for a system to be composed out of parts, so that the 
whole supervenience approach can get off the ground. I take this to be unproblematic here and say that a whole 
is composed if it has component parts. Using this notion of composition, 
holism is the claim that the whole has features that cannot be reduced to 
features of its component parts. Both \citet{healey91} and \citet{teller86} 
use the same kind of notion for the reduction relation, namely \emph{supervenience}. However, whereas Teller only
 speaks about relations of the whole and non-relational properties of the parts,
  Healey uses a broader view on what features of the whole should supervene on 
  what features of the parts. Because of its generality I take essentially 
  Healey's definition to be paradigmatic for the supervenience approach to 
holism\footnote{The exact definition by \citet[p.402]{healey91} is as follows.
`\emph{Pure physical holism}:
There is some set of physical objects from a domain $D$ subject
 only to processes of type $P$, not all of whose qualitative, intrinsic 
 physical properties and relations are supervenient upon the qualitative,
 intrinsic physical properties and relations of their basic physical parts 
 (relative to $D$ and $P$)'. The definition by \citet{teller86} is a
 restriction of this definition to solely relations of 
 the whole and intrinsic non-relational properties of the parts.
}\label{healey'sdef}. In this approach, holism in physical theories means 
that there are physical properties or relations of the whole  that are not
 supervenient on the intrinsic physical properties and relations of the 
 component parts. An essential feature of this approach is that the 
\emph{supervenience basis}, i.e., the properties or relations on which the whole 
 may or may not supervene,
are only the \emph{intrinsic} ones, which are those which the parts have at the time in question 
in and out of themselves, regardless of any other individuals.

We see that there are three different aspects involved in this approach.
The \emph{first} has to do with the metaphysical, or ontological effort of 
clarifying what it means that a whole is composed out of parts. I took this to be 
unproblematic. The \emph{second} aspect gives us the type of dependence the whole 
should have to the parts in order to be able to speak of holism. 
This was taken to be supervenience.
\emph{Thirdly}, and very importantly for the rest of this paper, the supervenience basis 
needs to be specified because  the supervenience criterion is 
 relativized to this basis. \citet[p.401]{healey91} takes this basis to be `just the qualitative, intrinsic
properties and relations of the parts, i.e., the properties and relations that
these bear in and out of themselves, without regard to any other objects, and
irrespective of any further consequences of their bearing these properties
for the properties of any wholes they might compose.' Similarly \citet[p.72]{teller86}
uses `properties \emph{internal} to a thing, properties which a thing has independently
of the existence or state of other objects.'

Although the choice of supervenience basis is open to debate because it is hard to specify precisely, the idea is that 
we should not add global properties or relations 
to this basis. It is supposed to contain
only what we intuitively think to be \emph{non-holistic}.
However, as I aim to show in the next sections,  
an alternative basis exists to which a criterion for holism 
can be relativized. This alternative basis, the \emph{resource basis} as I call it,
 arises when one adopts a different view when considering physical theories.
 For such theories allow not only for presenting us with an
 ontological picture of the world (although possibly only after an interpretation is provided),
  but also they allow for specific forms of property assignment and property
 determination. The idea then is that these latter processes, such as measurement or 
 classical communication, have 
 intuitively clear \emph{non-holistic features}, 
   which allow for an epistemological analysis 
   of whether or not a whole can be considered to be holistic or not.
   
However, before presenting this new approach, I discuss how the 
supervenience approach treats classical physics
and quantum mechanics (in the orthodox interpretation).  In treating these two theories I will first present 
some general aspects related to the structure of properties these theories allow 
for, since they are also needed in future sections.

\subsection{Classical physics in the supervenience approach}
\label{classical_sub}

Classical physics assigns two kinds of properties to a system.
State independent or fixed properties that remain unchanged (such as mass and charge)
 and dynamical properties associated with quantities called dynamical
  variables (such as position and momentum) \citep{healey91}. It is the
  latter we are concerned with in order to address holism in a theory
  since these are subject to the dynamical laws of the theory.
Thus in order to ask whether or not classical physics is holistic 
we need to specify how parts and wholes get assigned the dynamical properties in the
 theory\footnote{This presentation of the structure of properties in classical physics was inspired by \citet{isham95}.}. This \emph{ontological 
issue} is unproblematic in classical physics, for it views objects
as bearers of determinate properties (both fixed and dynamical ones). 
The \emph{epistemological issue} of how to gain knowledge of 
these properties is treated via the idea of \emph{measurement}. A measurement is any physical 
operation by which the value of a physical quantity can be inferred. 
Measurement reveals this value because it is assumed that the system has 
the property that the quantity in question has that value at the time of measurement. 
In classical physics there is no fundamental difference between 
measurement and any other
physical process. \citet[p.57]{isham95} puts it as follows: `Properties are intrinsically 
attached to the object as it exists in the world, and measurement is 
nothing more than a particular type of physical interaction designed 
to display the value of a specific quantity.' The bridge between ontology 
and epistemology, i.e., between property assignment (for any properties to 
exist at all (in the theory)) and property inference (to gain knowledge about them), 
is an easy and unproblematic one called measurement.

The specific way the dynamical properties of an object are encoded in the formalism of classical 
physics is in a state space $\Omega$ of physical states
 $x$ of a system. This is a phase space  where at each time a 
unique state $x$ can be assigned to the system. Systems or ensembles  can be 
 described by \emph{pure states} which are single points $x$ in $\Omega$ 
 or by  \emph{mixed states} which are unique convex combinations 
 of the pure states.
The set of dynamical properties determines the position of the system in the phase
 space $\Omega$ and conversely the dynamical properties of the system can
 be directly determined from the
 coordinates of the point in phase space. Thus, a \emph{one-to-one
 correspondence} exists between systems and their dynamical properties on the one hand, and 
 the mathematical representation in terms of points in phase space on the other.	 
 Furthermore, with observation of properties being unproblematic, the state corresponds 
 uniquely to the outcomes of the (ideal) measurements that 
 can be performed on the system.  The specific property 
 assignment rule for dynamical properties that captures the 
 above is the following.

A physical quantity $\mathfrak{A}$ is represented by a function
 $A:~\Omega \to  \mathbb{R}$
 such that $A(x)$ is the value $A$ possesses when the state is $x$. 
  To the property that the value of $A$ lies in the real-valued interval $\Delta$ there is associated
   a Borel-measurable subset
   \begin{equation}\label{1system}
    \Omega_{A\in\Delta}= A^{-1} \verb|{|\Delta\verb|}|=    \verb|{|x\in \Omega|A(x) \in \Delta\verb|}|,
\end{equation}
of states in $\Omega$ for which the proposition that the system 
has this property is true. Thus dynamical properties are associated with
\emph{subsets} of the space of states $\Omega$, and we have the 
one-to-one correspondence mentioned above between properties and points in the state space  now as follows: 
$ A(x)\in\Delta \Leftrightarrow x\in\Omega_{A\in\Delta} $.
Furthermore, the logical structure of the propositions about the dynamical properties 
of the system is identified with the \emph{Boolean $\sigma$-algebra} $\mathcal{B}$ of subsets of the space of
states $\Omega$. This encodes the normal logical way  (i.e., Boolean logic) of dealing with 
propositions about properties\footnote{The relation of
\emph{conjunction} of propositions corresponds to the 
 set-theoretical \emph{intersection} (of subsets of the state space),
  that of  \emph{entailment} between propositions to the 
 set-theoretical \emph{inclusion}, that of \emph{negation} of a proposition
  to the set-theoretical \emph{complement} and finally that of \emph{disjunction} of propositions
  corresponds to the set-theoretical \emph{union}. 
  In classical physics the (countable) logic of propositions
     about properties is thus isomorphic
    to a Boolean $\sigma$-algebra of subsets of the state space.
  }.

In order to address holism we need to be able to speak about 
properties of composite systems in terms of properties of the subsystems.
The first I will call \emph{global} properties, the second \emph{local}
 properties\footnote{Note that \emph{local} has here nothing to do with the issue of locality or spatial separation.
 It is taken to be opposed to global, i.e., restricted to a subsystem.}\label{local}.
It is a crucial and almost defining feature of the state 
space of classical physics that the local dynamical properties \emph{suffice} 
for inferring \emph{all} global dynamical properties. 
This is formalised as follows.
Consider the simplest case of a composite system with two subsystems (labeled $1$ and $2$). 
Let the tuple $<\Omega_{12},\mathcal{B}_{12}>$ characterize 
the state space of the composite system and the Boolean $\sigma$-algebra 
of subsets of that state space. The latter is isomorphic 
to the logic of propositions about the global properties. This tuple is determined 
by the subsystems in the following way. Given the tuples $<\Omega_1,\mathcal{B}_1>$
and $<\Omega_2,\mathcal{B}_2>$ that characterize the subsystem state spaces and 
property structures, $\Omega_{12}$ is the Cartesian product space of $\Omega_{1}$ and  $\Omega_{2}$,
i.e.,
\begin{equation} \label{cartesian}
\Omega_{12}=\Omega_1\times\Omega_2,
\end{equation}
 and furthermore,
\begin{equation}\label{algebras}
\mathcal{B}_{12}=\mathcal{A}(\mathcal{B}_1,\mathcal{B}_2),
\end{equation}
where $\mathcal{A}(\mathcal{B}_1,\mathcal{B}_2)$ is the 
smallest $\sigma$-algebra generated by $\sigma$-algebras that contain
 Cartesian products as elements. This algebra is defined by the following three
properties \citep{halmos88}: (i) if $\mathcal{A}_1\in\mathcal{B}_1$, $\mathcal{A}_2\in\mathcal{B}_2$ then
$\mathcal{A}_1\times\mathcal{A}_2 \in \mathcal{A}(\mathcal{B}_1,\mathcal{B}_2)$,
 (ii) it is closed under countable conjunction, disjunction and taking differences, 
(iii) it is the smallest one generated in this way. The $\sigma$-algebra
 $\mathcal{B}_{12}$ thus contains by definition all sets that can be written as a countable conjunction
 of Cartesian product sets such as
  $\Lambda_{1}\times\Lambda_{2}\subset\Omega_{12}$ 
  (with $\Lambda_{1}\subset\Omega_{1}$, $\Lambda_{2}\subset\Omega_{2}$), also called \emph{rectangles}.
 
The above means that the Boolean $\sigma$-algebra of the properties of the composite system 
is in fact the product algebra of the subsystem algebras.
Thus propositions about global properties (e.g.,  global quantity $B$ 
having a certain value) 
 can be written as disjunctions of 
propositions which are conjunctions of propositions about local properties alone
(e.g., subsystem quantities $A_1$ and $A_2$ having certain values). In other words,
the truth value of all propositions about $B$ can be determined from the truth value 
of disjunctions of properties about $A_1$ and $A_2$. The first and the latter thus have the same extension.

On the phase space $\Omega_{12}$ all this gives rise to the following structure.
To the property that the value of $B$ of a composite 
system lies in $\Delta$ there is associated a Borel-measurable subset of $\Omega_{12}$, for which the proposition that the system has this property
  is true:  
  \begin{equation}\label{samen}
    \{(x_{1}, x_{2})\in \Omega_{12}
|~B(x_{1},x_{2})\in \Delta\}\in\mathcal{B}_{12},
  \end{equation} 
 where $(x_{1},x_{2})$ are the pure states (i.e., points) in the phase space of the composite system
  and $x_1$ and $x_2$ are the subsystem states that each lie in the state space $\Omega_1$ or $\Omega_2$ of the respective subsystem.
  The important thing to note is that this subset lies in the product algebra $\mathcal{B}_{12}$
  and therefore is determined by the subsystem algebras $\mathcal{B}_{1}$ 
  and $\mathcal{B}_{2}$ via Eq. (\ref{algebras}).
  
From the above we conclude, 
and so is concluded in the supervenience approaches mentioned 
in the Introduction, although on other non-formal grounds,
that classical physics is not holistic. For the global properties supervene 
on the local ones because the Boolean algebra structure of the global properties is 
determined by the Boolean algebra structures of the local ones.
Thus all quantities pertaining to the global properties 
defined on the composite phase space such as $B(x_{1},x_{2})$ 
 are supervening quantities. 

For concreteness consider two examples of such supervening quantities $B(x_{1},x_{2})$
of a composite system. The first is $q= \parallel\vec{q_1}-\vec{q_2}\parallel$ which gives us the 
global property of a system that specifies the distance 
between two subsystems.
The second is $\overrightarrow{F}= -\vec{\nabla} V(\parallel \vec{q_1}-\vec{q_2}\parallel)$ 
which gives us the property of a system that says how strong the force is 
between its subsystems arising from the potential $V$. This could for example 
be the potential 
$\frac{m_1 m_2G}{\parallel\vec{q_1}-\vec{q_2}\parallel}$ for  the Newtonian gravity force.
Although both examples are highly non-local and could involve action at a distance, 
no holism is involved since the global
properties supervene on the local ones. As \citet[p.76]{teller86} puts it: 
`Neither action at a distance nor distant spatial separation
threaten to enter the picture to spoil the idea of the world
working as a giant mechanism, understandable in terms of the individual parts.'

Some words about the issue of whether spatial relations are to be considered
  holistic, are in order here.
Although the spatial relation of relative distance of the whole
indicates the way in which the parts are related with respect to position, 
whereby it is not the case that each of the parts has a position 
independent of the other one, it is here nevertheless not regarded a holistic property
since it is supervening on spatial position. We have seen that the distance $q$ between
two systems is treated supervenient on the systems having positions $\vec{q_1}$ 
and $\vec{q_2}$ in the sense expressed by Eq. (\ref{samen}).
 However, the argumentation given here 
requires an \emph{absolutist} account of space  so that position can be regarded as
 an intrinsic property of a system.
But one can deny this and adopt a \emph{relational} account of space 
and then spatial relations become monadic and positions become derivative,
which has the consequence that one has to incorporate spatial relations 
in the supervenience basis\footnote{A more subtle example than the relative distance between two points 
would be the question whether or not the relative angle between two 
directions at different points in space is a supervening property, 
i.e., whether or not the relative angle is to be considered 
  holistic or not. This depends on
  whether or not one can consider local orientations
 as properties that are to be included in the supervenience basis.}.\label{angle} 

On an absolutist account of space the spatial relation of relative distance 
  between the parts of a whole is shown to be supervenient upon local properties,
   and it is thus not to be included in the supervenience basis\footnote{\citet{teller87} for example takes spatial relations to be supervening on 
 intrinsic physical properties since for him the latter include spatiotemporal properties.}.
A relationist account, however, does include the spatial relations in the supervenience
 basis. The reason is that on this account they are to be regarded as 
\emph{intrinsically} relational, and therefore non-supervening on the subsystem properties. 
 \citet{cleland84} and \citet{french89} for example argue spatial relations
 to be non-supervening relations.  
 Furthermore, some hold that all other intrinsic relations can be regarded to be supervenient upon these.
The intuition is that wholes seem to be built out of their parts
 if arranged in the right spatial relations, and these spatial relations are taken 
 to be in some sense \emph{monadic} and therefore not holistic\footnote{\citet[p.409]{healey91} phrases this as follows: `Spatial relations are of special significance because 
they seem to yield the only clear example of qualitative, intrinsic
relations required in the supervenience basis in addition to the
 qualitative intrinsic properties of the relata. Other intrinsic
  relations supervene on spatial relations.'}.

Thus we see that issues depend on what view one has about the 
nature of space (or space-time).
 Here I will not argue for any position,
but merely note that if one takes an absolutist stance towards space
 so that bodies are considered to have a particular position, then spatial relations can be considered to be supervening 
on the positions of the relata in the manner indicated by the decomposition
of Eq. (\ref{samen}).  This discussion about whether spatial relations 
are to be regarded as properties that should be included in the supervenience basis 
clearly indicates that the supervenience criterion must be relativized to the supervenience basis.
As we will see later on this is analogous to the fact that the 
epistemological criterion proposed here work must be relativized to the resource basis.

As a final note in this section, I mention that because of the 
one-to-one correspondence in classical physics 
between physical quantities on the one hand and states on the state space on the other hand,
 and because composite 
states are uniquely determined by subsystem states (as can be seen
 from Eq.(\ref{cartesian})), it suffices to consider
the state space of a system to answer the question whether or not 
some theory is holistic. 
The supervenience basis is thus determined by the state space (supplemented with the fixed properties).
However, this is a special case and it contrasts with the quantum mechanical case 
(as will be shown in the next subsection). The supervenience approach should 
take this into account.
Nevertheless, the supervenience approach mostly limits itself
to the quantum mechanical state space in determining whether or not 
quantum mechanics is holistic.
The epistemological
approach to be developed here uses also other relevant features 
of the formalism, such as property determination, and focuses
therefore primarily on the structure of the assigned properties and not on that of the state space.
This will be discussed in the following sections.
 
\subsection{Quantum physics in the supervenience approach}
\label{quantum_sub}

In this section I will first treat some general aspects of the 
quantum mechanical formalism before discussing how the supervenience approach
deals with this theory.

In quantum mechanics, just as in classical physics, systems are assigned two kinds of properties. 
On the one hand, the fixed properties that we find in classical physics 
supplemented with some new ones such as intrinsic spin.
On the other hand, dynamical properties such as components of spin \citep{healey91}.
These dynamical properties are, again just as in classical physics, determined in a 
certain way by values observables have when the system is in a particular state.
 However, the state space and observables are
 represented quite differently from what we have already seen in
 classical physics. In general, a quantum state \emph{does not} correspond
  uniquely to the outcomes of the measurements that can be
 performed on the system. Instead, the system is assigned a specific
 Hilbert space $\mathcal{H}$ as its state space and the physical state of the system is represented by a state vector 
$\left | \, \psi \right \rangle$ in the pure case and a density operator $\rho$ in the mixed case.
Any physical quantity  $\mathfrak{A}$ is represented by an observable or
self-adjoint operator $\hat{A}$.
  Furthermore, the spectrum of $\hat{A}$ is the set of possible values the quantity 
   $\mathfrak{A}$ can have upon measurement.  
    
The pure state $\left | \, \psi \right \rangle$ 
     can be considered to assign a probability distribution 
          $p_i=| \left \langle \,\psi\right |  i\rangle|^2  $ to an orthonormal 
     set of states $\verb|{|\left | \, i \right \rangle\verb|}|$.
     In the case where one of the states is the vector $\left | \, \psi \right \rangle$, it is
      completely concentrated onto this vector. The state $\left | \, \psi \right \rangle$ can thus be regarded as the analogon of a
	     $\delta$-distribution on the classical phase space $\Omega$, 
        as used in statistical physics. However
	     the radical difference is that the pure quantum states
         do not (in general) form an orthonormal set. This implies that the
	     pure state $\left | \, \psi \right \rangle$ will also assign a positive probability to a
        different state  $\left | \, \phi \right \rangle$ if they are non-orthogonal and thus have 
        overlap. 
        This is contrary to the
	     classical  case, where the pure state $\delta(q-q_0, p-p_0)$ concentrated
	      on $(p_0,q_0) \in \Omega$ will always give rise to a probability
        distribution that assigns probability zero to every other pure state, since pure states on $\Omega$ 
        cannot have overlap.
	      Furthermore, the probability that the value of an observable $\hat{B}$ lies 
        in the real interval $X$ when the system is in the quantum state $\rho$ is 
        $Tr\,(\rho P_{\hat{ B},X })$ where $P_{\hat{B},X}$ is the projector associated to 
        the pair $(\hat{B},X)$ by the spectral theorem for self-adjoint operators. 
        This probability is in general not concentrated in $\{0,1\}$ 
         even when $\rho$ is a pure state.  Only in the special case that 
          the state is an eigenstate of the  observable $\hat{B}$ 
           is it concentrated in $\{0,1\}$, and the system is assigned 
           the corresponding eigenvalue with certainty. 
	     From this we see that there is no one-to-one correspondence 
       between values an observable can obtain and states of the quantum system.
       
Because of this failure of a one-to-one correspondence there are \emph{interpretations}
 of quantum mechanics that postulate \emph{different} connections between 
 the state of the system and the dynamical properties it possesses. 
 Whereas in classical physics this was taken to be unproblematic and natural, 
 in quantum mechanics it turns out to be problematic and non-trivial. 
 But a connection must be given in order to ask about any holism, 
 since we have to be able to speak about possessed properties and 
  thus an interpretation that gives us a property assignment rule is necessary.
 Here I will consider the well-known \emph{orthodox interpretation} of 
 quantum mechanics that uses the so called \emph{eigenstate-eigenvalue link} for this connection:
 a physical system has the property that quantity $\mathfrak{A}$ has a particular 
 value if and only if its state is an eigenstate of the operator $\hat{A}$
 corresponding to $\mathfrak{A}$. This value is the eigenvalue associated with
 the particular eigenvector. Furthermore, in the orthodox interpretation 
 measurements are taken to be ideal \emph{von Neumann measurements}, whereby  upon 
 measurement the system is projected into an eigenstate of the observable 
 being measured and the value found is the eigenvalue corresponding to 
 that particular eigenstate. The probability for this eigenvalue to
  occur is given by the well-known Born rule 
  $\left \langle \, i \right |\rho\left | \, i \right \rangle$, 
  with $\left | \, i \right \rangle$ the eigenstate that is projected upon and $\rho$ the state
   of the system before measurement.
Systems thus have properties \emph{only} if they are in an 
eigenstate of the corresponding observables, i.e., the system either already is or must first be projected into such an eigenstate
by the process of measurement. We thus see that the \emph{epistemological} 
scheme of how we gain knowledge of properties, i.e., the measurement process 
described above, serves also as an \emph{ontological} one defining what properties of a
 system can be regarded to exist at a given time at all.

Let me now go back to the supervenience approach to holism and ask what it says 
about quantum mechanics in the orthodox interpretation stated above.
According to all proponents of this approach mentioned 
in the Introduction quantum mechanics is holistic. The reason for this is supposed
 to be the feature of \emph{entanglement}, a feature absent in classical physics. In order to discuss
 the argument used, let me first present some aspects
  of entanglement. 
 Entanglement is a property of composite quantum systems whereby the state 
 of the system cannot be derived from any combination of the subsystem states.
 It is due to the tensor product structure of a composite 
 Hilbert space and the linear superposition principle of quantum mechanics.
In the simplest case of two subsystems, the precise definition is 
that the composite state $\rho$ cannot be written as a convex sum
 of products of  single particle states, i.e., $\rho \neq \sum_i p_i \rho^1_i\otimes\rho^2_i$, with $p_i\in[0,1]$ and $ \sum_ip_i=1$. 
In the pure case, an entangled state is one that cannot 
be written as a product of single particle states.
Examples include  the so-called \emph{Bell states} $\left | \, \psi^{-} \right \rangle$ and $\left | \, \phi^{-} \right \rangle$
 of a spin-$1/2$ particle. These states can be written as
 \begin{equation}\label{entang}
 \left | \, \psi^{-} \right \rangle=\frac{1}{\sqrt{2}}(\left | \, 01 \right \rangle_z -\left | \, 10 \right \rangle_z),~~~~~~
 \left | \, \phi^{-} \right \rangle=\frac{1}{\sqrt{2}}(\left | \, 00 \right \rangle_z -\left | \, 11 \right \rangle_z),
 \end{equation}
 with $\left | \, 0 \right \rangle_z$ and $\left | \, 1 \right \rangle_z$ eigenstates of the spin operator $\hat{S}_z=
\frac{\hbar}{2} \hat{\sigma}_z$, i.e., the spin up and down state in the $z$-direction respectively. These
 Bell states are eigenstates
  for total spin of the composite system given by 
the observable $\hat{S}^2=(\hat{S}_1+\hat{S}_2)^2$ with eigenvalue $0$ and $2\hbar^2$ 
  respectively. 
  
According to the orthodox
interpretation, if the composite system is in one of the states of 
Eq. (\ref{entang}),
the system possesses one of two global properties for total spin  which are completely different, namely 
eigenvalue $0$ and eigenvalue $2\hbar^2$. The question now is whether 
or not this spin property is holistic, i.e., does it or does it not supervene on subsystem properties? According to 
the supervenience approach it does not and the argument goes 
as follows. 
Since the individual subsystems have the same reduced 
state, namely the completely mixed state $\frac{1}{2}\mathds{1}$, and because these are not eigenstates of any spin observable, 
no spin property
at all can be assigned to them. So there is a difference in global properties to which no 
difference in the local properties of the subsystems corresponds. 
  Therefore there is no supervenience and we have an instantiation of
holism\footnote{This is the exact argument \citet{maudlin98} uses. 
 \citet{healey91} and \citet{esfeld01} also use an 
 entangled spin example whereas \citet{teller86,teller89}, \citet{french89} and \citet{howard89} use 
 different entangled states or some consequence of entanglement such as
  violation of the bipartite Bell inequalities.}.
  It is the feature of entanglement in this example that is 
  held responsible for holism. \citet{maudlin98} even defines
 holism in quantum mechanics in terms of entanglement and \citet[p.205]{esfeld01} puts it as 
  follows: `The entanglement of two or more states is the basis for the 
  discussion on holism in quantum physics.' Also \citet[p.11]{french89},
 although using a different approach to supervenience (see footnote \ref{voetnootfrech}), shares this view: 
`Since the state function [...] is not a product of the separate state functions 
of the particles, one cannot [...] ascribe to each particle an individual state 
function. It is {\em{this}}, of course, which reveals the peculiar non-classical
holism of quantum mechanics.'

      I would now like to make an observation of a crucial aspect of the reasoning
 the supervenience approach uses to conclude that quantum mechanics
  endorses holism. In the above and also in other cases the issue is treated via 
  the concept of entanglement of quantum states.
  This, however, is a notion primarily tied to the structure of the state space of 
  quantum mechanics, i.e., the Hilbert space, and not to the structure of the 
  properties assigned in the interpretation in question. There is no one-to-one 
  correspondence between states and assigned dynamical properties,
   contrary to what we have already seen in the classical case. Thus questions in terms of
   states, such as `is the state entangled?'  and in terms of properties such 
   as `is there non-supervenience?' are different \emph{in principle}. And although 
   there is some connection via the property assignment rule using the 
   eigenvalue-eigenstate link, I claim them to be relevantly different.
Holism is a thesis about the structure of properties assigned to a whole and to its parts,
not a thesis about the state space of a theory. 
The supervenience approach should carefully ensure that it takes this into account.
However, the epistemological approach of the next section naturally takes this into account 
since it focuses directly on property determination. It probes the structure
 of the assigned properties and  not just that of the state space.

The reason that in the supervenience approach one immediately and solely looks 
at the structure of the state space is because in its supervenience basis only 
the properties the subsystems have in and out of themselves at the time in 
question are regarded. 
This means that using the eigenstate-eigenvalue rule 
for the dynamical properties one focuses on properties the system has 
in so far as the state of the system implies them. Only eigenstates
give rise to properties, other states do not.
A different approach, still in the orthodox interpretation, would be to focus
 on properties the system can possess according 
 to the possible property determinations quantum mechanics allows for. It is the 
 structure of the properties that can be possibly assigned at all, which is then at the heart of 
 our investigations. In this view one could say that
 the physical state of a system is regarded more generally, as also 
 \citet{howard89} does, as a set of \emph{dispositions} for the system to manifest 
certain properties under certain (measurement) circumstances, whereby the 
eigenstates are a special case assigning properties with certainty.
This view is the one underlying the epistemological approach 
which will be proposed and worked out next.

\section{An epistemological criterion for holism in physical theories}
\label{criterion}

Before presenting the new criterion for holism I would like to motivate
it by going back to the spin-$1/2$ example of the last section.
Let us consider the example, which according to the supervenience 
approach gives an instantiation of holism, from a different point of view.  Instead of 
 solely considering state descriptions, let us look at what physical 
 processes can actually be performed according to the theory in question 
 in order to gain knowledge of the system. I call this an 
 \emph{epistemological stance}. I will show next that it then \emph{is} possible to 
 determine, using only non-holistic means (to be specified later on) whether 
 or not one is dealing with the Bell state  $\left | \, \psi^{-} \right \rangle$ or $\left | \, \phi^{-} \right \rangle$ of Eq. (\ref{entang}). How?
First measure on each subsystem the spin in the $z$-direction. 
Next, compare these results using classical communication.
 If the results have the same parity, the composite system was in the
  state $\left | \, \phi^{-} \right \rangle$  with global spin property $2\hbar^2$. And if the results do not have the same parity,
 the system was in the state  $\left | \, \psi^{-} \right \rangle$ with global spin property $0$.

 Thus using local measurements and classical communication the different global
 properties can be inferred after all.
   There is \emph{no} indication of holism in this approach, 
which is different from what the supervenience approach
  told us in the previous section.
 Although it remains true that the mixed reduced states of the individual 
 subsystems do not determine the composite state and neither 
 a local observable (of which there is no eigenstate), enough information 
 can be nevertheless gathered by local operations 
 and classical communication to infer the global property.
 We see that from an epistemological point of view we should not get 
 stuck on the fact that the subsystems themselves
  have no spin property because they are not in an eigenstate of a 
  spin observable. We can assign them a state, and thus 
  can perform measurements and assign them some local properties, which in this case do
   determine the global property in question.
 
From this example we see that this approach to holism  
does not merely look at the state space of a theory, but focuses on the structure of properties assigned to a whole and 
to its parts, as argued before that it should do. 
Then how do we spot \emph{candidates} 
for holism in this approach? Two elements are crucial.
Firstly, the theory must contain global properties that cannot 
be inferred from the local properties assigned to the subsystems,
 while, secondly, we must take into account 
 \emph{non-holistic constraints} on the  determination of these properties. 
 These constraints are that we only use the resource basis 
 available to local agents (who each have 
 access to one of the subsystems).  The guiding intuition is that 
 using this resource basis will provide us with only
 non-holistic features of the whole.  From this we finally get 
the following criterion for holism in a physical theory:
\begin{quote}
A physical theory is holistic 
if and only if it is impossible in principle, 
for a set of local agents each having access to a single subsystem only,
to infer the global properties of a system as assigned in the theory 
(which can be inferred by global measurements), 
by using the resource basis available to the agents.
\end{quote}
Crucial is the specification of the resource basis.
The idea is that these are all non-holistic resources for property 
determination available to an agent. 
    However, just as in the case of the specification of the supervenience basis, 
  this basis probably cannot be uniquely specified, i.e., the exact content of the 
  basis is open to debate. Here I propose that these resources include at least 
all \emph{local operations and classical communication} (abbreviated as LOCC)\footnote
{Note again that \emph{local} has here nothing to do with the issue of locality or 
spatial separation, but that it is taken to be opposed to global,
 i.e., restricted to a subsystem.}.
 The motivation for this is the intuition that local operations, i.e., anything we do on the separate subsystems, 
and classically communicating whatever we find out about it,
 will only provide us with  non-holistic properties of a composite system. 
However it could be possible to include other, although more debatable,
 non-holistic resources.
  A good example of such a debatable resource we have already seen: Namely,
 whether or not an agent can consider the position 
 of a subsystem as a property of the subsystem, so that he can 
 calculate relative distances when he knows the fixed positions of other subsystems.
 Another example is provided by the discussion of footnote \ref{angle}
 which suggests the question whether or not an agent can use a shared 
 Cartesian reference frame,  or a channel that transmits objects with well-defined
orientations, as a resource for determining the relative 
angle between directions at different points in space.

I believe that the determination of these and other spatial relations 
should be nevertheless included in the resource basis, for I take these 
relations to be (spatially) nonlocal, yet not holistic. Furthermore because we are dealing with epistemology
 in specifying the resource basis, I do not think that 
including them necessarily implies ontological commitment as to which view one
 must endorse  about space or space-time.
Therefore, when discussing different physical theories in the next section, 
I will use as the content of the resource basis, firstly, the determination 
of spatial relations, and secondly  LOCC (local operations and classical communication).
The latter can usually be unproblematically formalized within 
physical theories and do not depend on for example the view one has about 
spacetime. I thus propose to study the physical realizability 
of measuring or determining global properties while taking as a constraint
that one uses LOCC supplemented with the determination of spatial relations.

Let me mention some aspects of this proposed approach 
before it is applied in the next section.
\emph{Firstly}, it tries to formalize the question of holism in the 
context of what modern physical theories are, taking them 
 to be (i) schemes to find out and predict what the results are of
 certain interventions, which can be possibly used for determination of assigned properties,
  and (ii), although not relevant here, possibly describing 
  physical reality. Theories are no longer taken to necessarily present
  us with an ontological picture of the world specified by the
  properties of all things possessed at a given time.    
  
\emph{Secondly}, the approach treats the concept of \emph{property}
physically and not ontologically (or metaphysically). I mean by this that the concept is treated
analogous to  the way Einstein treated space and time (as that what is given by measuring rods
  and revolutions of clocks), namely as that which can be attributed to a system
  when measuring it, or as that which determines the outcomes of interventions.

\emph{Thirdly}, by including classical communication, this 
approach considers the possibility of determining some intrinsic relations among 
the parts such as the parity of a pair of bits, as was seen in the previous spin-{1/2} 
example. The parts are considered as parts, i.e., as constituting a whole
  with other parts and therefore being related to each other. 
   But the idea is that they are nevertheless considered non-holistically
    by using only the resource basis each agent has for determining 
    properties and relations of the parts. 

\emph{Fourthly}, as mentioned before,  the epistemological criterion for holism
 is relativized to the resource basis. Note that this is analogous to the 
 supervenience criterion which is relativized to the supervenience basis.
 I believe this relativizing to be unavoidable and even desirable because 
 it, reflects the ambiguity and debatable aspect inherent in any discussion 
 about holism. Yet, in this way it is incorporated in a fair and clear way. 

 \emph{Lastly}, note that the epistemological criterion is logically
independent of the supervenience criterion. Thus whether or not a theory is holistic in the supervenience approach
 is independent of whether or not it is holistic in the 
 newly proposed epistemological approach. This is the case because not all 
 intrinsic properties and relations in the supervenience basis are necessarily 
 accessible using the resource basis, and conversely, some that are accessible using the resource 
 basis may not be included in the supervenience 
 one\footnote{Of the latter case an example was given using the spin 
1/2 example, since the property specifying whether the singlet state or the 
triplet state obtain is not supervening, but can be inferred using only 
LOCC. Of the first case an example will be given in the next section.}.

\section{Holism in classical physics and quantum mechanics; revisited. }
\label{QMsection}
In this section I will apply the epistemological criterion for holism 
to different physical theories, where I use as the content of the resource basis 
the determination of spatial relations supplemented with LOCC.

\subsection{Classical Physics and Bohmian Mechanics }
\label{classbohm}

In section \ref{classical_sub} classical physics on a phase space was deemed non-holistic 
in the supervenience approach because global properties in this 
theory were argued to be supervening on subsystem properties.
Using the epistemological criterion we again find that 
classical physics is deemed non-holistic\footnote{Note that in both cases only systems with finite many subsystems are considered.}. 
The reason is that because of the one-to-one relationship between
properties and the state space and the fact that a Cartesian product is
 used for combining subsystem state spaces,
 and because measurement in classical physics 
is unproblematic as a property determining process, the resource basis 
allows for determination of all subsystem properties. We thus
are able to infer the Boolean $\sigma$-algebra of the properties
of the subsystems. Finally, given this the global 
properties can be inferred from the local ones  (see section \ref{classical_sub}), because the Boolean algebra structure 
of the global properties is determined by the Boolean algebra structures of the local ones,
 as was given in Eq. (\ref{algebras}). Hence no epistemological holism can be found.

Another interesting theory that also uses a state space with 
a Cartesian product to combine state spaces of subsystems
 is \emph{Bohmian mechanics} (see e.g. \citet{durr95}). 
 It is not a phase space but a configuration space.
This theory has an ontology of 
particles with well defined positions on 
trajectories\footnote{\emph{Bohmian mechanics}, which has as 
ontologically existing only particles with well defined positions on 
trajectories, should be distinguished (although this is perhaps 
not common practice) from  the so-called the \emph{de Broglie-Bohm theory} 
 where besides particles also the wave function has ontological existence as a guiding field. 
 This contrasts with Bohmian mechanics since in this theory the wave function has 
 only nomological existence.  Whether or not de Broglie-Bohm theory is 
 holistic because of the different role assigned to the wave function 
   needs careful examination, which will here not be executed.
  }. Here I discuss the interpretation where this theory is supplemented with 
   a property assignment rule just as in 
classical physics (i.e., all functions on the state space correspond
 to possible properties that can all be measured). Indeed, pure physical 
 states of a system are given by single points $(\vec{q})$ 
 of the position variables $\vec{q}$  that together make up a configuration space.
  There is a one-to-one relationship between the set of properties a system has 
 and the state on the configuration space it is in, 
as was shown in section \ref{classical_sub}. 
    The dynamics is given by the 
possibly non-local quantum potential $U_{QM}(\vec{q})$ determined by
  the quantum mechanical state
  $\left | \, \psi \right \rangle$, supplemented with the ordinary classical potential 
 $V(\vec{q})$, such that the force on a particle is given by: 
 $\vec{F}\equiv\frac{d\vec{p}}{dt}=-\vec{\nabla}[V(\vec{q})+ U_{QM}(\vec{q})]$.
This theory can be considered to be a real mechanics, i.e., 
a Hamilton-Jacobi theory, although with a specific extra interaction term. 
This is the quantum potential in which 
 the wave function appears that has only nomological existence. (Although a
 Hamilton-Jacobi theory, it is not classical mechanics: the latter is a second order
  theory, whereas Bohmian mechanics is of first order, i.e.,
   velocity is not independent of position).
  
 In section \ref{classical_sub} all theories on a state space
 with a Cartesian product to combine subsystem state  spaces and
  using a property assignment rule just as in classical physics 
  were deemed  non-holistic by the supervenience approach and therefore we can conclude that 
 Bohmian mechanics is non-holistic in this approach.
 Perhaps not surprising, but the epistemological approach also 
deems this theory non-holistic. The reason why is the same as why classical physics 
as formulated on a phase space was argued above to be not holistic in this approach.

Because Bohmian mechanics and quantum mechanics in the orthodox interpretation
have the same empirical content, one might think that because the first is not holistic, neither is the latter.
However, this is not the case, as will be shown next. This illustrates the fact that an interpretation of a 
theory, in so far as a property assignment rule is to be given, 
is \emph{crucial} for the question of holism. A formalism on its own is not enough.

\subsection{Quantum Operations and Holism}
\label{qmholism}

In this section I will show that quantum mechanics in the orthodox 
interpretation is holistic using the epistemological criterion,
without using the feature of entanglement.
 In order to do this we need to specify  what the resource basis looks like
 in this theory. Thus we need to  formalize what a local operation is and what 
 is meant by classical communication  in the context of quantum mechanics. 
For the argument it is not necessary to deal
 with the determination of spatial relations and these will thus not be considered. 

  Let us first look  at a general quantum process $\mathcal{S}$ that takes 
  a state $\rho$ of a system  on a certain Hilbert space $\mathcal{H}_1$ to a different 
  state $\sigma$ on a possibly different Hilbert space $\mathcal{H}_2$, i.e., 
\begin{equation}
\rho \rightarrow \sigma=\mathcal{S} (\rho) ,~~~~~~~~~~\rho \in \mathcal{H}_1,~~ \mathcal{S} (\rho)\in\mathcal{H}_2,
\end{equation} 
where $\mathcal{S}: \mathcal{H}_1\rightarrow\mathcal{H}_2$ is a \emph{completely positive 
trace-nonincreasing map}. This is  an operator $\mathcal{S}$, positive and trace non-increasing,
 acting linearly on Hermitian matrices such that $\mathcal{S}\otimes\mathds{1}$ takes 
 states to states. These maps are  also called \emph{quantum operations}\footnote{See \citet{nielsen00} for an introduction to the general formalism of
 quantum operations.}.  Any quantum process, such as for example unitary evolution or measurement, 
  can be represented by such a quantum operation.

We are now in the position to specify the class of LOCC operations. It is the 
class of local operations plus two-way classical communication. It
consists of compositions of elementary operations 
of the following two forms
\begin{equation}
\mathcal{S}^A\otimes\mathds{1},~~~~~ \mathds{1}\otimes\mathcal{S}^{B},
\end{equation}
with $\mathcal{S}^A$ and $\mathcal{S}^B$ arbitrary local quantum operations.
 The class contains the identity and is closed under composition and taking tensor products.
As an example consider the case where  $A$ performs a measurement and communicates 
her result $\alpha$ to $B$, after which $B$ performs his measurement:
\begin{equation}
\mathcal{S}^{AB}(\rho)=(\mathds{1}\otimes\mathcal{S}^{B}_{\alpha})\circ(\mathcal{S}^A\otimes\mathds{1})(\rho).
\end{equation} 
We see that $B$ can \emph{condition} his measurement on the outcome that $A$ obtained.
 This example can be extended to many such rounds in which $A$ and $B$ 
each perform certain local operations on their part of the system and condition their choices on what is communicated to them.

Suppose now that we have a physical quantity $\mathfrak{R}$ of a bi-partite system
with a corresponding operator $\hat{R}$ that has 
a set of nine eigenstates, $\left | \,\psi_1 \right \rangle$ to $\left | \,\psi_9 \right \rangle$, with eigenvalues $1$ to $9$.
The property assignment we consider is the following:
if the system is in an eigenstate $\left | \, \psi_{i} \right \rangle$
then it has the property that quantity $\mathfrak{R}$ has the fixed value $i$ 
(this is the eigenstate-eigenvalue link).
Suppose $\hat{R}$ works on $\mathcal{H}=\mathcal{H}_A\otimes\mathcal{H}_B$ (each three dimensions) and
 has the following complete orthonormal set of \emph{non-entangled} eigenstates:
\begin{eqnarray}\label{productstates}
\left | \,\psi_1 \right \rangle &=& \left | \, 1 \right \rangle\otimes\left | \, 1 \right \rangle,\nonumber\\
\left | \,\psi_{2,3} \right \rangle&=& \left | \, 0 \right \rangle\otimes\left | \, 0\pm1 \right \rangle,\nonumber\\
\left | \,\psi_{4,5} \right \rangle&=& \left | \, 2 \right \rangle\otimes\left | \, 1\pm2 \right \rangle,\nonumber\\
\left | \,\psi_{6,7} \right \rangle&=& \left | \, 1\pm2 \right \rangle\otimes\left | \, 0 \right \rangle,\nonumber\\
\left | \,\psi_{8,9} \right \rangle&=& \left | \, 0\pm1 \right \rangle\otimes\left | \, 2 \right \rangle,
\end{eqnarray}
with $\left | \, 0+1 \right \rangle=\frac{1}{\sqrt{2}}(\left | \, 0 \right \rangle+\left | \, 1 \right \rangle)$, etc.

We want to infer whether the composite system has the property that the value
of the observable $\mathfrak{R}$ is one of the numbers $1$ to $9$, using 
only LOCC operations performed by two observers $A$ and $B$, who each have 
one of the individual subsystems.
Because the eigenstate-eigenvalue link
is the property assignment rule used here,
 this amounts to determining which eigenstate $A$ and $B$
 have or project on during the LOCC measurement.
If $A$ and $B$ project on eigenstate $\left | \, \psi_{i} \right \rangle$ 
then a quantum operation
 $\mathcal{S}_i:\rho \rightarrow \frac{S_i(\rho)}{{\mathrm{Tr}}
 [S_i(\rho)]}$ is associated to the measurement outcome $i$, with projection 
 operators $S_i=\left | \, i \right \rangle_A\left | \, i \right \rangle_B
 \left \langle \, \psi_i \right |$. This 
 is nothing but the well-known projection due to measurement 
(with additional renormalisation), but here written in the 
 language of local quantum operations\footnote{Instead of writing the projection operators as  
$S_i=\left | \, \psi_{i} \right \rangle \left \langle \, \psi_i \right |$, I write  $S_i=\left | \, i \right \rangle_A\left | \, i \right \rangle_B
\left \langle \, \psi_i \right |$ 
to show explicitly that \emph{only} local records are taken. Since the states $\left | \, i \right \rangle$ 
can be regarded eigenstates of some local observable, we
  can regard them to determine a local property using
   the property assignment rule in terms
    of the eigenvalue-eigenstate link of the orthodox interpretation.} .
   The state  $\left | \, i \right \rangle_A$ denotes 
 the classical record of the outcome of the measurement 
that $A$ writes down, and similarly for $\left | \, i \right \rangle_B$.
These records can be considered to be \emph{local} properties of the subsystems 
$A$ and $B$.

It follows from the theory of quantum operations \citep{nielsen00} that 
implementing the quantum operation $\mathcal{S}(\rho)=\sum_i\mathcal{S}_i\rho\mathcal{S}_i^{\dagger}$
amounts to determining the \emph{global} property assignment given by $\hat{R}$. 
Surprisingly, this cannot be done using LOCC, a result obtained by \citet{bennett99}.
For the complete proof see the original article by \citet{bennett99} or 
\citet{walgate02}\footnote{This result is a special case of the fact that 
some family of separable quantum operations (that all have a complete eigenbasis of separable states)
cannot be implemented by LOCC and von Neumann measurements. 
This is  proven by \citet{chen03}.}, but a sketch of it goes as follows. 
If $A$ or $B$ perform von Neumann measurements in any of 
 their operation and communication rounds then the distinguishability of the 
 states is spoiled. Spoiling occurs in \emph{any} local basis. The ensemble 
 of states as seen by $A$ or by $B$ alone is therefore non-orthogonal,
  although the composite states are in fact orthogonal.

  From this we see that a physical quantity, whose corresponding operator has only product 
eigenstates, gives a property assignment using the eigenvalue-eigenstate link that
is not measurable using LOCC. Furthermore, we see that the resource basis sketched before does not 
suffice in determining the global property assignment given by $\hat{R}$.
 Thus according to the epistemological criterion of the previous section  
quantum mechanics is holistic, although no entanglement is involved.
 Examples of epistemological holism that do involve entanglement can of course be given. 
 For example, distinguishing the four (entangled) Bell states given by 
 $\left | \, \psi^{+} \right \rangle$, $\left | \, \psi^{-} \right \rangle$, $\left | \, \phi^{+} \right \rangle$ and $\left | \, \phi^{-} \right \rangle$
  (see Eq.(\ref{entang})) cannot be done by LOCC. 
  Thus entanglement is \emph{sufficient} to prove epistemological holism. However,
  this is hardly surprising. What is surprising is the fact that it 
  is \emph{not necessary}, i.e., that here a proof 
  of epistemological holism is given not involving entanglement.
  Furthermore because of the lack of entanglement in this example 
 it would not allow for a proof of holism in the supervenience approach.
  Of course, it may well be that the resource basis used in this example is too limited,
 but I do not see other resources that may sensibly be included in this basis 
 so as to render this example epistemologically non-holistic.

\section{Conclusion and outlook}
\label{conc}

I sketched an epistemological criterion for holism that determines,
once the resource basis has been specified, whether or not a physical
theory with a property assignment rule is holistic.
It was argued to be a suitable one for addressing holism in physical theories,
because it focuses on property determination
as specified by the physical theory in question 
(possibly equipped with a property assignment rule via an interpretation).
 I distinguished this criterion from the well-known supervenience criterion for
holism and showed them to be logically independent.
Furthermore, it was shown that both the epistemological and the supervenience
approaches require relativizing the criteria to respectively 
the resource basis and the supervenience basis. I argued that in general 
neither of these bases is determined by the state space of a physical theory. In other words, 
 holism is not a thesis about the state space a theory uses, 
 it is about the structure of properties and property 
 assignments to a whole and its parts that a theory or an interpretation 
 allows for.  And in investigating what it allows for 
 we need to try to formalize what we intuitively 
 think of as holistic and non-holistic.  
  Here, I hope to have given a satisfactory new epistemological
 formulation of this, that allows one to 
 go out into the world of physics and apply the new criterion to the theories or 
 interpretations one encounters.

In this paper I have only treated some specific physical theories.
It was shown that all theories on a state space using a Cartesian 
product to combine subsystem state spaces, 
 such as classical physics and Bohmian mechanics, are not holistic 
 in both the supervenience and epistemological approach.
  The reason for this is that the Boolean algebra structure of the global properties is 
determined by the Boolean algebra structures of the local ones.
 The orthodox interpretation of quantum mechanics, however, was found 
 to instantiate holism. This holds in both approaches,
 although on different grounds. For the supervenience approach it 
 is the feature of entanglement
 that leads to holism, whereas using only LOCC resources, one can have
epistemological holism in absence of any entanglement, i.e., 
when there is no holism according to the supervenience approach.
  
There are of course many open problems left.   What is it that we can 
single out to be the reason of the holism found?
The use of a Hilbert space with its feature of superposition? Perhaps, but not the kind of superposition
 that gives rise to entanglement, for I have argued that it is not entanglement
  that we should per se consider
 to be the paradigmatic example of holism.  Should we blame the property assignment rule which 
 the orthodox interpretation uses? I shall leave this an open problem.

The entangled Bell states $\left | \, \psi^{-} \right \rangle$ and $\left | \, \phi^{-} \right \rangle$  
of section \ref{quantum_sub} could,
 despite their entanglement, be distinguished after all using 
only LOCC, whereas this was not possible in the set of nine (non-entangled)
product states  of Eq.(\ref{productstates}). These two quantum mechanical examples show us that 
we can do both more and less than quantum states at first seem to tell us. 
This is an insight gained from the new field of \emph{quantum information theory}.
 Its focus on what one \emph{can} or \emph{cannot do} do with quantum 
systems, although often from an engineering point of view, has produced a
 new and powerful way of dealing with questions in the foundations of quantum 
 mechanics that can lead to fundamental new insights or principles. 
 I hope the new criterion for holism in physical theories suggested 
 in this  paper is an inspiring example of this.

\section*{Acknowledgments}I would like to thank Jos Uffink and Victor Gijsbers for stimulating discussions 
and valuable remarks. Furthermore, I would like to thank an anonymous referee for
 very helpful suggestions to reformulate the thesis defended here 
 and to strengthen the argumentation considerably.

\end{document}